# OPTICAL SETI OBSERVATIONS OF THE ANOMALOUS STAR KIC 8462852

[SHORT TITLE: OPTICAL SETI OBSERVATIONS OF KIC 8462852]

Marlin Schuetz[1,2], Douglas A. Vakoch[1*], Seth Shostak[3], Jon Richards[3]

[1]SETI International, 100 Pine St., Ste. 1250, San Francisco, CA 94111-5235, USA
[2]Boquete Optical SETI Observatory, Volcancito Road, Boquete, Chiriquí 0413, Panama
[3]Center for SETI Research, SETI Institute, 189 Bernardo Ave., Mountain View, CA 94043, USA

**Abstract**

To explore the hypothesis that KIC 8462852's aperiodic dimming is caused by artificial megastructures in orbit (Wright et al. 2015), rather than a natural cause such as cometary fragments in a highly elliptical orbit (Marengo et al. 2015), we searched for electromagnetic signals from KIC 8462852 indicative of extraterrestrial intelligence. The primary observations were in the visible optical regime using the Boquete Optical SETI Observatory in Panama. In addition, as a preparatory exercise for the possible future detection of a candidate signal (Heidmann 1991), three of six observing runs simultaneously searched radio frequencies at the Allen Telescope Array in California. No periodic optical signals greater than 67 photons/m$^2$ within a time frame of 25 ns were seen. This limit corresponds to isotropic optical pulses of 8 10$^{22}$ joules. If, however, any inhabitants of KIC 8462852 were targeting our solar system (Shostak & Villard 2004), the required energy would be reduced greatly. The limits on narrowband radio signals were 180 – 300 Jy Hz at 1 and 8 GHz, respectively, corresponding to a transmitter with an effective isotropic radiated power of 4 10$^{15}$ W (and 7 10$^{15}$ W) at the distance of KIC 8462852. While these powers requirements are high, even modest targeting could – just as for optical signals – lower these numbers substantially.

## 1. Introduction

KIC 8462852, a star system examined by NASA's Kepler telescope, has received significant attention due to its unusual light curve. Whereas stars with exoplanets exhibit slight periodic dimming due to transits of exoplanets, KIC 8462852's dimming is aperiodic and of much greater magnitude—up to 22% of the stellar flux.

---

[*] Corresponding author. Email: dvakoch@setiinternational.org

Numerous explanations have been proposed for this anomalous behavior. While a catastrophic collision in an asteroid belt of KIC 8462852 might explain the dimming, such an impact would produce dust particles that would absorb starlight and reradiate at lower frequencies, yielding an infrared excess. However, no such infrared excess has been detected either before or after these anomalous dimming events (Marengo et al. 2015). A more plausible explanation is that dimming is due to cometary fragments in a highly elliptical orbit, which would not cause excess infrared radiation (Boyajian et al. 2015, Marengo et al. 2015). Nevertheless, even this explanation needs further investigation to determine the mechanism for such putative cometary fragmentation.

An alternative explanation for the non-periodic dimming of KIC 8462852 posits the existence of artificial megastructures in orbit around the star (Wright et al. 2015). A Dyson swarm in orbit could capture KIC 8462852's starlight for a civilization's energy use, although depending on the total surface area of the solar satellites, this might also produce an infrared excess (Dyson 1960). In addition to a Dyson swarm, it's possible to imagine other artificial structures that might account for the dimming of starlight but result in only a negligible increase in infrared emission, for example structures made of highly reflective films. Engineered objects the size of planets could serve as long-term beacons detectable through their transits with the Kepler or CoRoT missions (Arnold 2005, Arnold 2013). In the current study we ask whether a civilization capable of such a large-scale engineering project might also be transmitting intentional electromagnetic signals. An earlier search for extraterrestrial intelligence (SETI) using the Allen Telescope Array (ATA) in northern California was conducted for narrowband and broadband radio signals between 1 GHz and 10 GHz (Harp et al. 2015). No signals were detected.

The current experiment expanded this search to a hunt for brief laser pulses at optical wavelengths using the Boquete Optical SETI Observatory in Panama, a small, privately owned facility that has been active in optical SETI searches since 2010. Located at an altitude of 1300 m on the slopes of a dormant volcano near the Costa Rican border, the observatory's principal instrument is a 0.5 m Newtonian telescope outfitted with a single photomultiplier detector. In addition, as a preparatory exercise for efforts that would be desirable in case of the actual detection of a candidate signal (Billingham et al. 1991; Heidmann 1991), a subset of observing runs simultaneously searched both optical and radio frequencies at the Boquete observatory and the ATA.

**2. Detecting Brief Laser Pulses Against a Stellar Background**

An advanced technological civilization seeking to signal evidence of its existence at interstellar distances might do so using a powerful laser with pulse duration in the nanosecond range (Schwartz & Townes 1961; Howard & Horowitz 2001; Howard et al. 2004). Stars with known exoplanets have been targeted in past SETI searches (Siemion et al. 2013), with exoplanets discovered by the Kepler mission being of particular interest because any electromagnetic communications between planets would be along an orbital plane viewed edge-on from the perspective of our solar system, making it conceivable we could detect this spillover radiation (Tellis & Marcy 2015). Several papers over the past decade have described megajoule (MJ)-scale pulsed lasers as being capable of significantly outshining the parent star at distances up to approximately 1000 light years. Specifically, during the period of a pulse there could be several orders of magnitude greater photon flux produced by the laser than by the star. For example, at a distance of 100 light-years, a Sun-like star will produce a steady flux of $10^8$ photons/sec-m$^2$, or about 0.1 photons/nanosecond. In that same interval, a laser pulse could cause tens or even thousands of photons/m$^2$—a disparity that provides a clear path for the design of detectors.

The method most frequently used for signal detection in optical SETI employs multiple photomultipliers arranged to work as a coincidence detector. These have been needed to discriminate against large amplitude signals commonly produced in photomultipliers by gamma radiation, corona effects, and Cerenkov radiation. To avoid such interfering phenomena, it is common practice to employ optical beam splitter(s) to feed multiple photomultipliers or avalanche photodiodes. When each of the detectors registers photon(s) that are coincident in time, an output is produced. To manage the level of false positives it has been usual to set the discriminators' threshold levels for two or more photoelectron pulses. Note that increasing the threshold levels is done at the expense of reducing the sensitivity.

For the small telescope at the Boquete Optical SETI Observatory a different approach was needed to minimize the sensitivity trade-off and to avoid the losses and complications with beam splitters, etc. The solution was twofold: a purpose built photometer having a single photomultiplier and fast Fourier transform processing of the photometer pulsed output.

In this scheme, the photomultiplier output pulses follow two paths. The first is to a high-level discriminator whose threshold is set to detect two or more coincident photoelectron pulses. The other path has a low-level discriminator to eliminate photomultiplier dark noise, followed by an integrator/discriminator whose threshold can be set to detect groups of photoelectron pulses spread out in time, in the present case a ~25 ns interval. The group discriminator threshold can be set to detect 2, 3 or more photoelectron pulses in the interval. The choice for this selection is dependent upon the stellar background level. (Non-coincident photoelectron pulses are monitored because there may be technical and economic reasons to use lesser peak power lasers with longer duration pulses. Alternatively the senders could use several even lower peak power lasers, operating at different wavelengths, and fired sequentially to convey a message).

Of course the single photomultiplier method has inherently more residual noise, which without additional processing would render the technique only marginally useful. However, we mitigate this limitation with additional processing as follows: When a coincident or group event occurs, the photometer sends a pulse to a computer-based low frequency fast Fourier transform spectrum analyzer. Analysis of very short duty cycle pulsed signals by this method has a particularly beneficial consequence. When a periodic signal is detected and processed, it does not produce a single major transform peak as expected with a sinusoidal signal (Smith 1997). It displays instead the fundamental frequency and all of the harmonics at nearly equal amplitudes. Thus, a precisely periodic signal has an unmistakable signature, and it can be detected even when the stellar background flux is very large. False positives are extremely unlikely.

The detection scheme is routinely tested at the photometer using simulated signals produced by two internal LEDs. One LED is used to simulate stellar backgrounds up to about 750,000 cps. A second LED is pulsed at an arbitrary 0.6 Hz and at a very low power level, i.e., ~1 microwatt at 2 – 50 ns pulse durations. The light emitted from both LEDs must undergo dispersive and attenuating reflections before passing through a 0.7 mm photometer aperture. In this manner, the light impinging upon the detector photocathode can be controlled to the individual detected photon level. Among other routine tests it is usual to power the pulsed LED while raising the simulated stellar background levels to a point where the signal to background counts approach unity. Those tests have demonstrated that for 2 detected photoelectron pulses (n=2) in the 25 ns interval, the processed pulsed signal is 10 dB above a 20 kcps background. With the threshold set for 3 detected photoelectron pulses (n=3) the processed pulsed signal is 10 dB above a 200 kcps background. Of course the photometer discriminator threshold can easily be set higher for bright stars, but in order to maintain the highest sensitivities to laser signals, the majority of observations at the Boquete observatory have been limited to stars having visual magnitudes ≥7 and more preferably ≥9. It is also noteworthy that the pulsed LED power level can be set low enough so as to produce only group events. Raising the power level slightly can demonstrate coincident events in addition to group events.

### 3. Optical Observations and Results

Optical SETI observations were conducted in October and November 2015 at the Boquete observatory in Panama. Observation periods of up to 1 hour were limited by the position and westerly advance of the star. Weather conditions limited the useful observations to October 29 and November 9, 10, 23, 24, and 28 (UT).

A photometer mounted on the telescope employed a single high gain photomultiplier sensitive in the 300 - 600 nm range. Detected photons, in the form of ~3 ns FWHM pulses are processed in the photometer with circuitry that may be adjusted to create an output pulse whenever 2, 3, or more photons are detected within any time window of ~25 ns. That includes coincident photon events (photon pileup) as well as group photon events where the photons are spread out in the time window. The two classes of events were monitored separately. For these observations and depending upon the sky-plus-stellar background, the discriminator was set to respond to either 2 or 3 detected photons within the period. Considering the telescope's aperture area and the various losses in the system, 2 detected photons correspond to a photon flux of ~67 photons m$^{-2}$.

For KIC 8462852, a star of visual magnitude 11.7, the stellar background count was 1500 to 2000 cps depending upon the atmospheric conditions. A computer-based fast Fourier transform spectrum analyzer was used to process the photometer output pulses in near real time in an attempt to detect any pulsed signal having a periodicity of between 0.05 and 10 Hz. (A signal at a repetition rate greater than 10 Hz, i.e., up to 100 Hz, would be easily detected, but was not expected. Detection of signals below 0.05 Hz would require additional processing and/or other observational techniques perhaps involving two or more observatories).

No periodic signals were detected. Considering the large distance to KIC 8462852 (454 pc) and the small telescope aperture, it is still possible that a pulsed laser signal, even if directed towards us, could be well below the 67 photons $m^{-2}$ detection limit of the Boquete instrument. However, it has been suggested that an advanced civilization might use very precise targeting to produce much larger photon fluxes on the basis of observed transits of the Sun by the Earth (Shostak & Villard 2004). For example, a 5 megajoule, 450 nm laser pulse beamed from KIC 8462852 toward Earth and diverging to an approximate 1.5 AU diameter disk would result in a signal detectable at the Boquete observatory.

**4. Simultaneous Radio Observations and Results**

Most SETI searches are conducted at a single observatory, without simultaneous efforts at other facilities. The existing post-detection SETI protocol indicates that in the event of a detection of extraterrestrial intelligence, astronomers should coordinate observations of the target at multiple sites (Billingham et al. 1991). The global SETI network that Heidmann (1991) envisioned would initially be restricted to the radio domain, reflecting the dominance of that regime in SETI a quarter century ago, but it could be expanded to include optical SETI today. In the event of a detection of an actual signal from extraterrestrial intelligence, it would be valuable to observe the target simultaneously across multiple regimes to characterize more comprehensively the full range of any signals being transmitted. Prior coordination between observatories would facilitate rapid response, which would be especially critical for relatively transient signals. Past efforts to conduct multifrequency, multisite SETI programs have been limited by including observatories both with and without SETI-specific signal processing capabilities (Narusawa et al. 2011) or by allowing different facilities either to observe different targets or to observe the same target at different times (Narusawa et al. 2013).

As a test of the feasibility of simultaneously observing the same target in optical and radio regimes, the ATA reobserved KIC 8462852 at a subset of the frequencies previously reported (Harp et al. 2015) on three of the six nights during which optical observations were conducted. The ATA's fully automated system for spectral analysis searched for narrowband emissions between 0.01 and 10 Hz wide, with real-time follow-up of any candidate signals (Tarter et al. 2011). Because the frequency of narrowband signals could change over time due to relative accelerations caused by diurnal rotation and orbital motions, the ATA's signal detection algorithms allow for detection of a fractional drift rate in frequency up to $10^{-9}$ /sec (1 Hz/sec at 1 GHz to 10 Hz/sec at 10 GHz).

Frequencies between 1699 and 1787 MHz were examined on October 29, with this full frequency range repeated three times. Observations on November 10 were conducted between 1400 and 1488 MHz, with each frequency examined twice. On November 28, frequencies between 8000 and 8200 MHz were examined once. Each frequency observation lasted 92 seconds, with the detection threshold equal to 6.5 times the mean square noise level. This yielded a sensitivity of 180 and 300 Jy Hz at 1 and 8 GHz, respectively, corresponding to a transmitter with an effective isotropic radiated power of $4 \cdot 10^{15}$ W (and $7 \cdot 10^{15}$ W) at the distance of KIC 8462852. No persistent narrowband signals were detected in the vicinity of KIC 8462852, consistent with previous results with the same instrument (Harp et al. 2015).

## 5. Conclusions

We have conducted a search for brief laser pulses from the vicinity of KIC 8462852 whose anomalous dimming has caused some to suggest the star may be orbited by an artificial megastructure. Our study was an attempt to detect any intentional laser signals transmitted from an extraterrestrial civilization in the KIC 8462852 system. No periodic optical signals greater than 67 photons/m$^2$ within a time frame of 25 ns were seen.

This limit corresponds to isotropic optical pulses of $8 \cdot 10^{22}$ joules. We can compare this to the National Ignition Facility's laser constellation, which can generate nanosecond pulses of several $10^6$ joules. Largely due to the distance between Earth and KIC 8462852, the energy demands for a detectable signal are significantly greater than in our terrestrial example.

If, however, any inhabitants of KIC 8462852 were targeting our solar system (Shostak & Villard 2004), the required energy would be reduced greatly. As an example, if such hypothetical extraterrestrials used a 10 m mirror to beam laser pulses in our direction, then using a 10 m receiving telescope, the minimum detectable energy per pulse would be 125,000 joules. If this pulse repeated every 20 minutes, then the average power cost to the transmitting civilization would be a rather low 100 watts. This would be a negligible cost for any civilization capable of constructing a megastructure large enough to be responsible for the dimming seen with KIC 8462852, particularly if that structure were used to capture a large fraction of the star's energy ($\sim 10^{27}$ watts). It would be considerably easier to detect such signals intentionally directed toward Earth than to intercept collimated communications between two star systems along a vector that accidentally intersects the Earth (Forgan 2014).

While the power requirement for a detectable, isotropic narrowband radio transmission from KIC 8462852 is obviously quite high (Harp et al. 2015), even modest targeting on the part of the putative extraterrestrials can reduce these numbers substantially. If we assume that the extraterrestrials are using an antenna the same size as the Arecibo radio telescope in Puerto Rico, the minimum detectable transmitter power (again, for narrowband signals) is only about 200 megawatts for frequencies near 1 GHz, and 4 megawatts for frequencies close to 10 GHz. These are, quite obviously, easily manageable power levels for any advanced society.

The method used for optical pulse detection in this study has some significant advantages over the traditional methods using multiple photomultipliers. Specifically, a detector designed to take advantage of the Poisson distribution at group numbers between 2 and 4 requires (1) only a single photomultiplier, (2) works equally well with coincidence and group pulse detection, (3) has simpler setup and calibration and no concern for channel crosstalk, (4) eliminates signal losses related to beam splitters, (5) suppresses random stellar background noise to a low level, (6) simplifies detector cooling (if needed), (7) can be adjusted according to the total count, e.g., $n=2$ for dim stars and $n=3$ or $4$ for brighter stars, (8) benefits from the fact that a longer pulse width requires less laser peak power for a given pulse energy, and (9) with more energy per pulse, but with the same laser peak power, more photons per pulse may be available for detection.

The use of longer pulse lengths could yield significant advantages. Consider the case of a 1 MJ laser with a 1 ns pulse with a peak power of 1 petawatt. If the pulse were lengthened to 50 ns, the peak power would be reduced by a factor of 50 for the same energy expenditure and total photon flux per pulse. The use of lasers transmitting the same energy but with longer pulses and with less peak power can reduce the facility costs, as well as resources and maintenance costs. If there are technical or economic reasons to do so, multiple lower power lasers might be employed to fire sequentially—filling out long pulse lengths and perhaps having different wavelengths for encoding data. It is also reasonable to expect that lower power lasers can function with greater energy efficiency. For both reception and transmission the use of longer pulse lengths appears to be a win-win strategy.

A civilization attempting to make contact could use a diverse range of transmission strategies, including, for example, serially targeting thousands or millions of stars. It is therefore plausible that pulsed emissions directed toward target stars could be at a very low rate, i.e., <<0.01 pps. In that event seemingly non-periodic pulses could be detected using synchronized data collection with two or more observatories. Once detected, observation times can be extended to look for periodicity. To increase opportunities for such collaboration between facilities, SETI International has begun developing a network of optical SETI observatories, with early members including the Boquete Optical SETI Observatory and the Owl Observatory in Michigan, USA (Howard 2015).